\documentclass{article}
\usepackage{spconf,amsmath,graphicx}
\usepackage[colorlinks,citecolor=blue,urlcolor=blue,bookmarks=false,hypertexnames=true]{hyperref}

\usepackage{enumitem}
\setlist{nosep, leftmargin=14pt}

\usepackage{mwe} 
\usepackage{amssymb}   

\title{Unpaired MRI Super Resolution with Contrastive Learning}
\name{Hao Li$^{1,\dagger}$, Quanwei Liu$^{2,\dagger}$, Jianan Liu$^{3,\star}$, Xiling Liu$^{4}$, Yanni Dong$^{5,\star}$, Tao Huang$^{2}$, and Zhihan Lv$^{6}$
\thanks{$^{\dagger}$ The first two authors contributed equally to this work.}
\thanks{$^{\star}$ Corresponding authors.}
}

\address{$^{1}$ The Department of Neuroradiology, University Hospital Heidelberg, Heidelberg, Germany; \\
$^{2}$ The College of Science and Engineering, James Cook University, Cairns, Australia;\\
$^{3}$ Vitalent Consulting, Gothenburg, Sweden; \\
$^{4}$ Department of Stomatology, Shenzhen People’s Hospital, 
Shenzhen, China; \\
$^{5}$School of Resource and Environmental Sciences, Wuhan University, Wuhan, China; \\
$^{6}$ Department of Game Design, Faculty of Arts, Uppsala University, Sweden.
}

\begin{document}
%
\maketitle
\begin{abstract}
Magnetic resonance imaging (MRI) is crucial for enhancing diagnostic accuracy in clinical settings. However, the inherent long scan time of MRI restricts its widespread applicability. Deep learning-based image super-resolution (SR) methods exhibit promise in improving MRI resolution without additional cost. Due to lacking of aligned high-resolution (HR) and low-resolution (LR) MRI image pairs, unsupervised approaches are widely adopted for SR reconstruction with unpaired MRI images. However, these methods still require a substantial number of HR MRI images for training, which can be difficult to acquire. To this end, we propose an unpaired MRI SR approach that employs contrastive learning to enhance SR performance with limited HR training data. 
Empirical results presented in this study underscore significant enhancements in the peak signal-to-noise ratio and structural similarity index, even when a paucity of HR images is available. These findings accentuate the potential of our approach in addressing the challenge of limited HR training data, thereby contributing to the advancement of MRI in clinical applications.
\end{abstract}
\section{Introduction}
\label{sec:intro}

Magnetic Resonance Imaging (MRI) is widely used for diagnosis and monitoring of treatment progress non-invasively \cite{li2021review}. However, the high cost and long scan time make it challenging to acquire high-resolution (HR) MRI images and limit the use of MRI in areas such as surgical guidance. To address this issue, single-image super-resolution (SR) techniques have been developed using deep learning-based approaches to enhance image resolution directly from low-resolution (LR) images \cite{zhao2020smore,wu2023assured,chen2018brain}.

Typically, these techniques rely on convolutional neural networks (CNNs) to learn the transformation from LR images to HR images. Since super-resolution reconstruction (SRR) is a complex and challenging problem, various powerful CNN architectures have been developed to improve the effectiveness of SR \cite{zhao2019channel,li20213d}. These supervised learning (SL) methods require a large number of paired HR and LR images. However, patient movements are always inevitable, leading to geometric distortions of soft tissues. As a result, acquiring aligned pairs of MRI images is extremely difficult. To solve this problem, unsupervised learning (UL) methods are adopted for MRI SRR. UL SR does not require pairs of images and achieves comparable results to supervised models in real-world image SRR \cite{zhu2017unpaired,yuan2018unsupervised,Wei_2021_CVPR,Maeda_2020_CVPR}. Inspired by the real-world image UL SRR methods, unsupervised MRI SR methods have been developed in response to recent advances in the field and eliminate the demand for paired training images\cite{9999498,liu2022unsupervised}.

However, these unsupervised methods still require a significant number of HR magnetic resonance images for training \cite{liu2022unsupervised,zhou2023unaen}. To reduce the dependence on a large amount of HR MRI images, contrastive learning (CL) \cite{he2020momentum, chen2020simple,wang2021unsupervised} can be used to extract supervisory information from a small amount of unpaired MRI images and train the network with this self-generated supervisory information. This approach allows contrastive-based models to identify meaningful representations for unsupervised or weakly supervised MRI SR tasks.

\begin{figure*}[htb]
  \centering
  \centerline{\includegraphics[width=14cm]{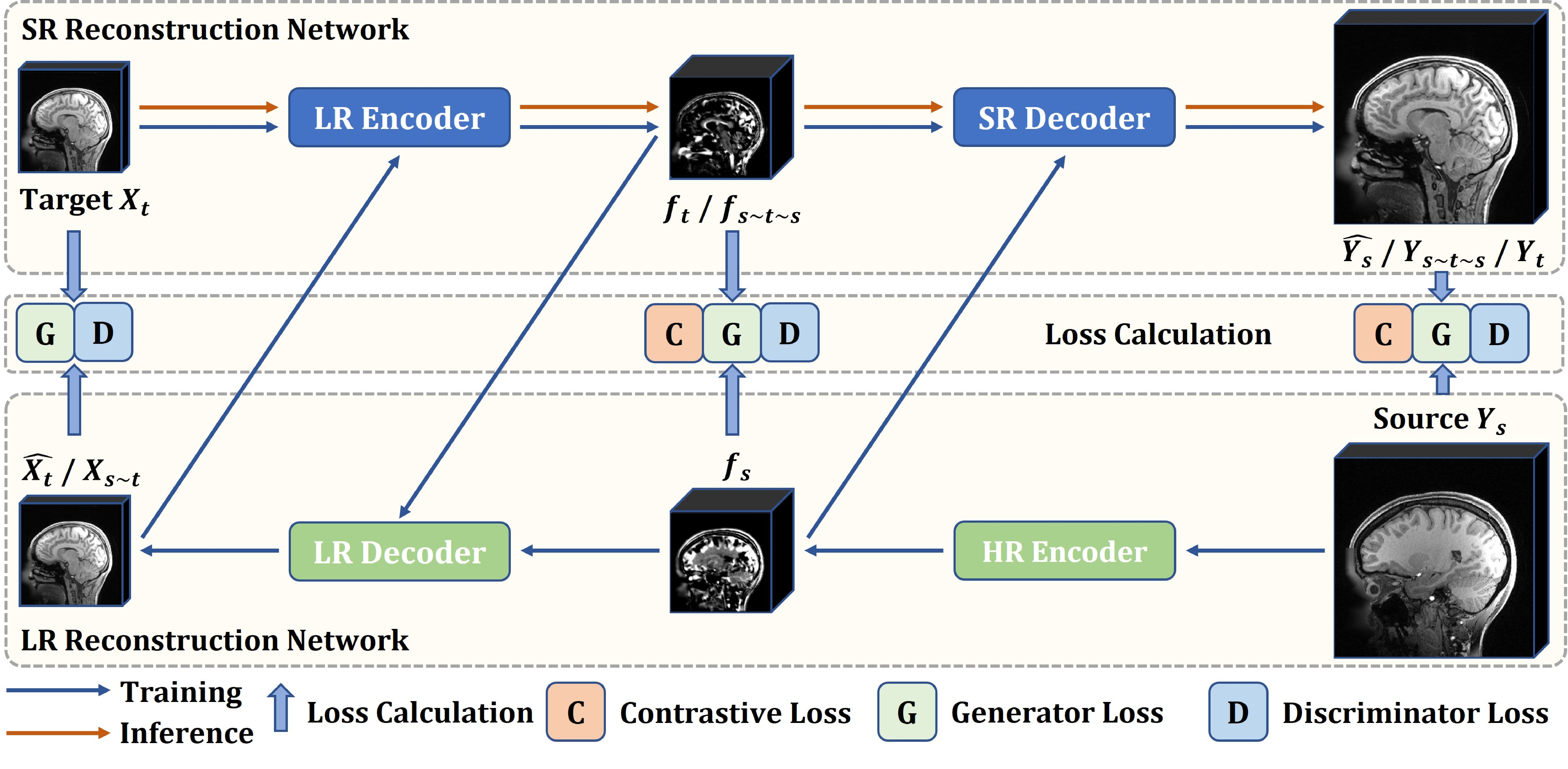}}
  \vspace{-0.4cm}
\caption{The diagram of the proposed unsupervised contrastive learning model with unpaired LR-HR MRI images. \textbf{C}, \textbf{G}, \textbf{D} are contrastive loss, generator loss and discriminator loss, respectively. The target $X_t$ and source $Y_s$ are fed into the model together to train the model, followed by three data flows: $Y_s \to f_s \to \widehat{Y_s}$, $Y_s \to f_s \to X_{s\sim t} \to f_{s\sim t\sim s} \to Y_{s\sim t\sim s}$, and $X_t \to f_t \to \widehat{X_t}$. Contrastive loss and generator losses are calculated between the data with the same content, such as $f_{s\sim t\sim s}$ to $f_s$ and $\widehat{Y_s}/Y_{s\sim t\sim s}$ to $Y_s$. The discriminator loss is calculated between the data in the same domain but with different contents, such as $X_{s\sim t}$ to $X_t$ and $f_t$ to $f_s$. For inference, only $X_t$ is fed to the model and follows the data flow of $X_t \to f_t \to \widehat{Y_t}$. }

\label{fig:res1}
\end{figure*}

This study presents an unpaired MRI SR framework that utilizes CL. The framework integrates InfoNCE \cite{he2020momentum} loss into an unsupervised SR architecture. Our contributions can be summarized as follows:
\begin{itemize}
  \item We present an efficient strategy grounded in CL for unpaired MRI SRR. This strategy involves the construction of positive sample pairs between generated SR images and the corresponding HR images, while negative samples are fashioned by pairing the generated SR images with other HR images in the same batch.
  \item We devise a contrastive representative learning-based unsupervised network, which furnishes a novel and effective approach to unpaired MRI SR. Compared to various state-of-the-art unsupervised approaches, our proposed model yields competitive outcomes.
  \item The performance of the proposed contrastive learning method with a limited number of HR training data is comparable to that of unsupervised learning methods with a large number of training data. Furthermore, when only a limited number of training data is provided, the unsupervised learning methods perform worse than ours, demonstrating the superiority of our method.
\end{itemize}

\section{PROPOSED METHOD}
\label{sec:format}
\subsection{Contrastive Learning and Sample Generation}
\label{sssec:subsubhead1}
Contrastive learning has emerged as a prominent framework for acquiring invariant feature representations of data. Previous research has extensively explored image feature extraction via CL, achieving performance that is comparable to or even surpassing supervised learning in diverse high-level tasks, such as classification, semantic segmentation, and object detection \cite{he2020momentum,chen2020simple,grill2020bootstrap,chen2021exploring}. However, the conventional CL pipeline is less suitable for low-level vision tasks \cite{wu2023practical}.

Traditional CL faces difficulties with low-level vision tasks such as SR. The process involves creating pairs of positive and negative samples to reduce the distance from predicted results to positive samples, while increasing the separation to negative ones. However, this becomes challenging when using data augmentation to generate sequences of positive and negative sample pairs, as it can lead to issues with maintaining dense pixel correspondences. As a result, the effectiveness of the traditional CL approach is diminished. Despite this, a few studies have explored the application of CL in real-world SR \cite{luo2022deep}. For example, Wang et al. \cite{wang2021towards} used negative samples from other datasets to train their network, while CRL-SR \cite{zhang2021blind} used CL to extract resolution-invariant features and recover lost or corrupted high-frequency details.

Our research introduces a simple strategy to construct positive and negative sample pairs. As depicted in Fig. \ref{fig:res1}, we incorporate a feature contrastive loss and a reconstruction contrastive loss established between the generated feature maps (or SR images) and the extracted feature map (or source HR images). CL is implemented using a contrastive loss, with one of the classical variants being infoNCE \cite{he2020momentum}. The work \cite{zhu2020deep} modified infoNCE by introducing inter- and intra-embedding terms in graph learning. For the ${i}$-th image, the loss is expressed as:
\begin{equation}\label{equation 2}
{\cal L}({u_i},{v_i}) = \log \frac{{{e^{\theta ({u_i},{v_i})/\tau }}}}{{{e^{\theta ({u_i},{v_i})/\tau }} + {\cal L}_{te}({u_i},{v_k})  + {\cal L}_{tr}({u_i},{u_k}) }},
\end{equation}
\begin{equation}\label{equation 3}
{\cal L}_{te}({u_i},{v_k}) =  \sum\limits_{k = 1}^N \mathbb{I}{{_{[k \ne i]}}{e^{\theta ({u_i},{v_k})/\tau }}},
\end{equation}
\begin{equation}\label{equation 4}
{\cal L}_{tr}({u_i},{u_k}) = \sum\limits_{k = 1}^N \mathbb{I}{{_{[k \ne i]}}{e^{\theta ({u_i},{u_k})/\tau }}}.
\end{equation}

Here  ${u_i}$ and ${v_i}$ represent the nonlinear embeddings generated from training samples. $\theta ({u_i},{v_i}) $  signifies the cosine similarity between ${u_i}$ and ${v_i}$, which is equivalent to the dot product between L2-normalized  ${u_i}$ and ${v_i}$ , e. g.  $\theta ({u_i},{v_i}) = u_i^T\cdot{v_i}/||u_i^T||\cdot||{v_i}||$ . The temperature parameter $\tau$ governs the impact of penalties on hard negative samples. $ \mathbb{I}_{[k \ne i]} $ acts as an indicator function. $N$ denotes the batch size. ${\cal L}_{te}({u_i},{v_k})$ and ${\cal L}_{tr}({u_i},{u_k})$ pertain to inter- and intra-embedding calculations, respectively. Since the two embeddings are symmetric, the final optimization goal is to minimize the average overall positive pairs:
\begin{equation}\label{equation 5}
{\cal L}_{cl} = -\frac{1}{{2N}}\sum\limits_{i = 1}^N {[{\cal L}({u_i}, {v_i}) + {\cal L}({v_i}, {u_i})]}.
\end{equation}
This formula can also be applied to CL involving images. Due to the separate calculation method, this approach is more memory-saving.  Therefore, this form of contrastive loss is adopted in our study.

\subsection{Unpaired MRI SR Architecture }

The proposed model is mainly derived from UDEAN \cite{liu2022unsupervised}, which is the recently proposed unsupervised approach for unpaired MRI SRR, and further incorporates CL. It comprises two main components: the representation generation module and the loss calculation module, as illustrated in Fig. \ref{fig:res1}.

The representation generation module consists of two key elements: the LR reconstruction network (depicted in green) and the SRR network (depicted in blue). The LR reconstruction network initially encodes the source domain dataset $Y_s$ into the feature space. Subsequently, the LR decoder network generates the LR image from the feature map $f_s$. In contrast, the objective of the SRR network is to extract features from the target domain dataset $X_t$ and decode them into an SR image $Y_t$.

In our approach, the LR encoder consists of six identical convolution modules, each comprising a convolution layer followed by an activation layer using the LeakyReLU function. The HR encoder shares a similar structure with the LR encoder, except for the initial convolutional layer, which employs a stride of 2 to downsample the HR MRI image to match the size of the LR feature map. We utilize a modified RCAN network \cite{li20213d,zhang2018image,lin2022revisiting} architecture as the backbone of the decoders (SR and LR decoders). This choice not only enhances computational efficiency but also ensures superior inference performance.

The loss calculation module consists of three components: the discrimination loss module (represented by \textbf{D}), the generator loss module (represented by \textbf{G}), and the contrastive loss module (represented by \textbf{C}). The discrimination loss module employs three discriminators, each of which is responsible for determining the authenticity of the source domain, the feature space, and the target domain. In particular, domain adaptation is integrated into this process, significantly improving the quality of the reconstructed SR image. For the discrimination network, we employ three VGG networks. The generator loss module incorporates various reconstruction losses, including the L1 loss and the structural similarity (SSIM) loss, to guarantee the fidelity of the generated SR images. Lastly, the contrastive loss module enhances the model's feature extraction capability by increasing the separation between samples within the batch and reducing the distance between the generated samples and the ground truth.

To apply contrastive learning in the proposed model, the positive sample pairs are generated in both feature space ($f_s $ and  $f_{s\sim t\sim s} $) and image space ($Y_s$ and $Y_{s\sim t\sim s}$, $Y_s$ and $\widehat{Y_t}$), which are expected to have the same content. And the negative samples are constructed by pairing the generated SR images and feature maps with other HR images and their feature maps within the same batch.


\section{EXPERIMENTS}
\label{sec:pagestyle}

\subsection{Data preparation}
\label{ssec:subhead1}

In this study, our dataset is derived from T1w images sourced from the Human Connectome Project (HCP) dataset which comprises 1113 participants. We conducted experiments on a randomly selected subset of 300 participants. This subset was divided into different groups, including the source group (120 participants), the target group (120 participants), the validation group (30 participants) and the evaluation group (30 participants). The four groups were isolated from each other. To assess the performance of the model and its dependence on varying amounts of HR samples, we further curated new source groups by selecting 70\%, 50\%, 30\% and 10\% of participants from the source group. The target group was down-sampled using 3D K-space truncation with a scale factor of 2×2×2. A scale factor of 1×2×2 appears to be optimal since maintaining the frequency encoding steps does not lead to additional acquisition time. However, most MRI scanners do not allow different resolutions in the frequency and phase encoding directions without modifying the pulse sequences. Therefore, the scale factor of 2×2×2 was used in this study.


\begin{figure*}[htb]
  \centering
  \centerline{\includegraphics[width=\textwidth]{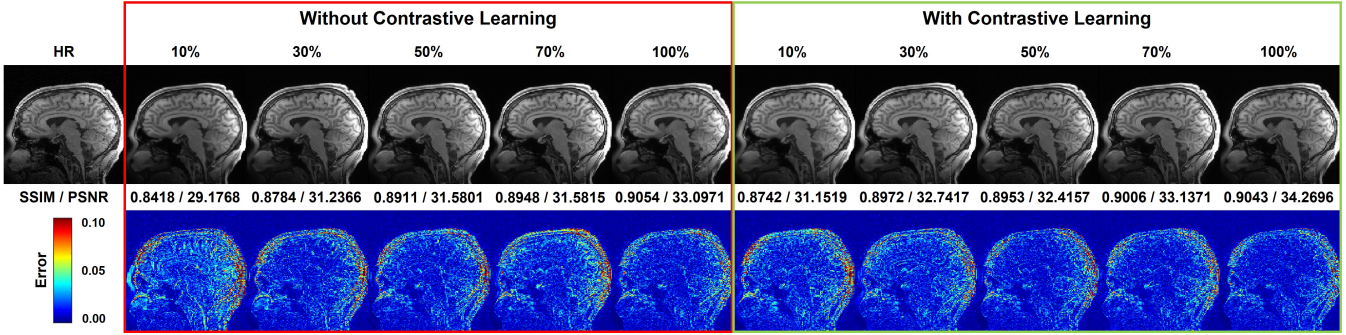}}
  \vspace{-0.2cm}
\caption{Comparison in visual effect and error maps with various numbers of training HR images. The visualization shows the super-resolution image in the sagittal plane of the HCP dataset which is downsampled with a scale factor of 2×2×2. }
\label{fig:res2}
\end{figure*}

\subsection{Implementation Details}
\label{ssec:subhead2}

To expedite the training process of our model, we utilized the distributed data-parallel method, training the model on two NVIDIA GeForce RTX 3090 GPUs. Our deep learning framework included PyTorch 1.9 and Lightning 2.0. For stable training, we set the learning rate of the discriminator to 0.00005 and that of the generator to 0.0002. We employed two well-established metrics: the peak signal-to-noise ratio (PSNR) and the structure similarity index (SSIM) to evaluate image quality.

\subsection{Results and Analysis}
\label{ssec:subhead3}
\subsubsection{Comparison with state-of-the-art UL MRI SR models}
\label{sssec:subsubhead}

We employed several state-of-the-art UL methods for comparison, including ZSSR \cite{Shocher_2018_CVPR}, DASR \cite{Maeda_2020_CVPR}, Pseudo SR \cite{Wei_2021_CVPR}, Blind-SR \cite{9999498}, and UDEAN \cite{liu2022unsupervised}. All of these methods utilized the modified RCAN network \cite{li20213d,zhang2018image,lin2022revisiting} as their underlying architecture, adopted hyperparameter settings of UDEAN, and were trained with 100\% HR training data to ensure equitable comparisons.

Table \ref{metrics} shows the numerical results, demonstrating that our proposed model outperforms the comparison models in nearly all cases. Although the SSIM of our model is slightly lower than that of UDEAN (by 0.0036), a substantial PSNR improvement of 0.7292 dB is observed.

\begin{table}
    \centering
\footnotesize    \caption{Quantitative comparison with other unsupervised SRR methods on HCP dataset with 100\% HR training data.}
    \label{metrics}
    \begin{tabular}{c|c|cc}
    \hline
     {Model name}& \multicolumn{1}{c|}{Learning} & \multicolumn{1}{c}{SSIM }&\multicolumn{1}{c}{PSNR} \\
     \hline
     Tricubic &  -& 0.8981 ± 0.0106 & 31.5862 ± 1.8520 \\
     ZSSR\cite{Shocher_2018_CVPR} & UL & 0.8994 ± 0.0163 & 33.0213 ± 2.2054 \\
     DASR\cite{Maeda_2020_CVPR} & UL & 0.8931 ± 0.0094 & 32.3309 ± 1.4671 \\
     Pseudo SR \cite{Wei_2021_CVPR} & UL & 0.8931 ± 0.0116 & 31.0407 ± 1.8816 \\
     Blind-SR \cite{9999498} & UL & 0.9132 ± 0.0100 & 33.0695 ± 1.4656  \\
     UDEAN \cite{liu2022unsupervised}   & UL & \textbf{0.9231\textbf{ ± }0.0083}  & 33.2484 ± 1.8029 \\
     \hline
     Ours & UL+CL & 0.9195 ± 0.0077  & \textbf{33.9776 ± 1.7226}  \\
     \hline
    \end{tabular}
\end{table}

\subsubsection{Comparison of different numbers of HR training images}

\begin{figure}[htb]

\begin{minipage}[b]{0.49\linewidth}
  \centering
  \centerline{\includegraphics[width=4.8cm]{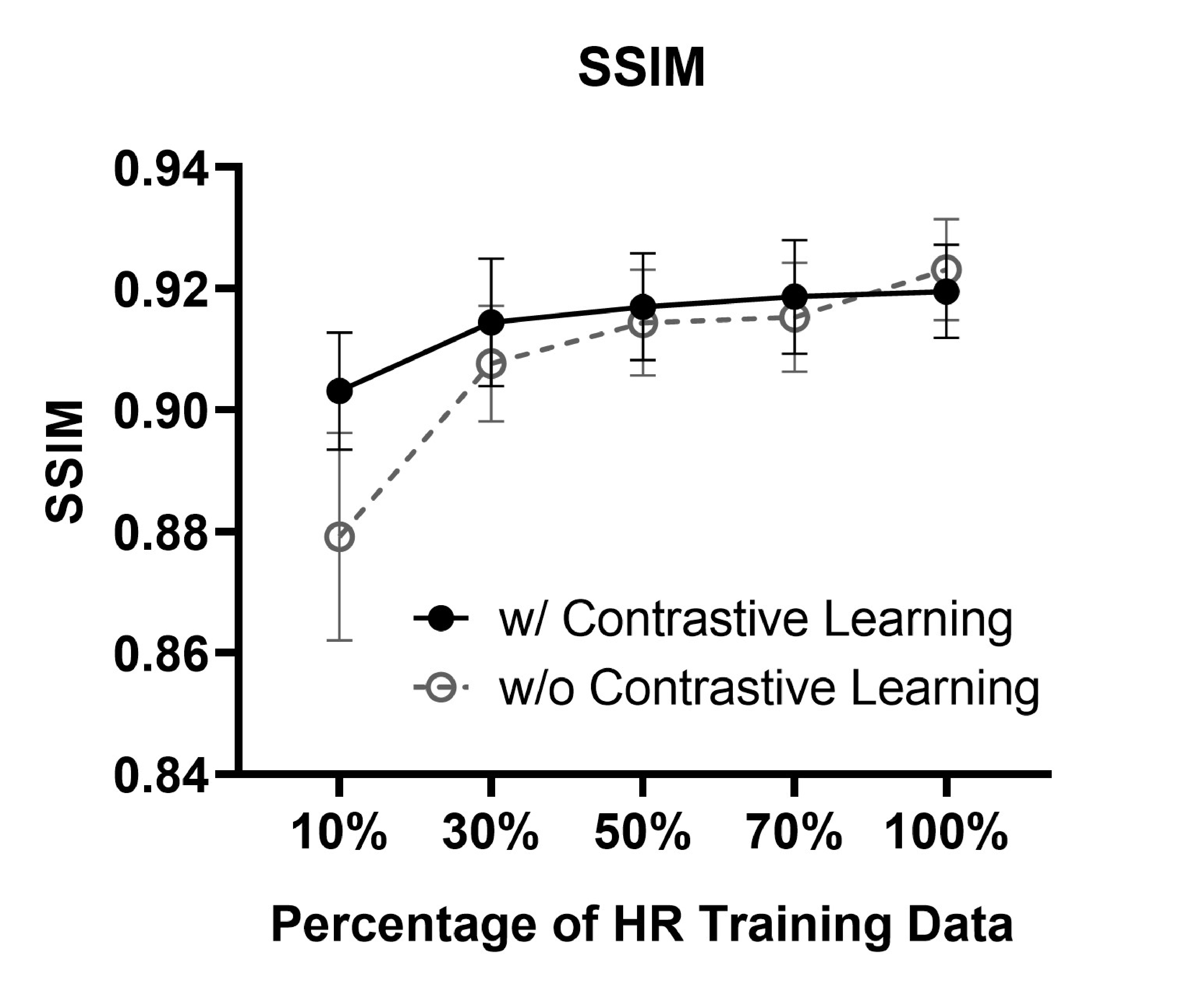}}
  \vspace{-0.1cm}
  \centerline{(a) }\medskip
\end{minipage}
\hfill
\begin{minipage}[b]{0.49\linewidth}
  \centering
  \centerline{\includegraphics[width=4.8cm]{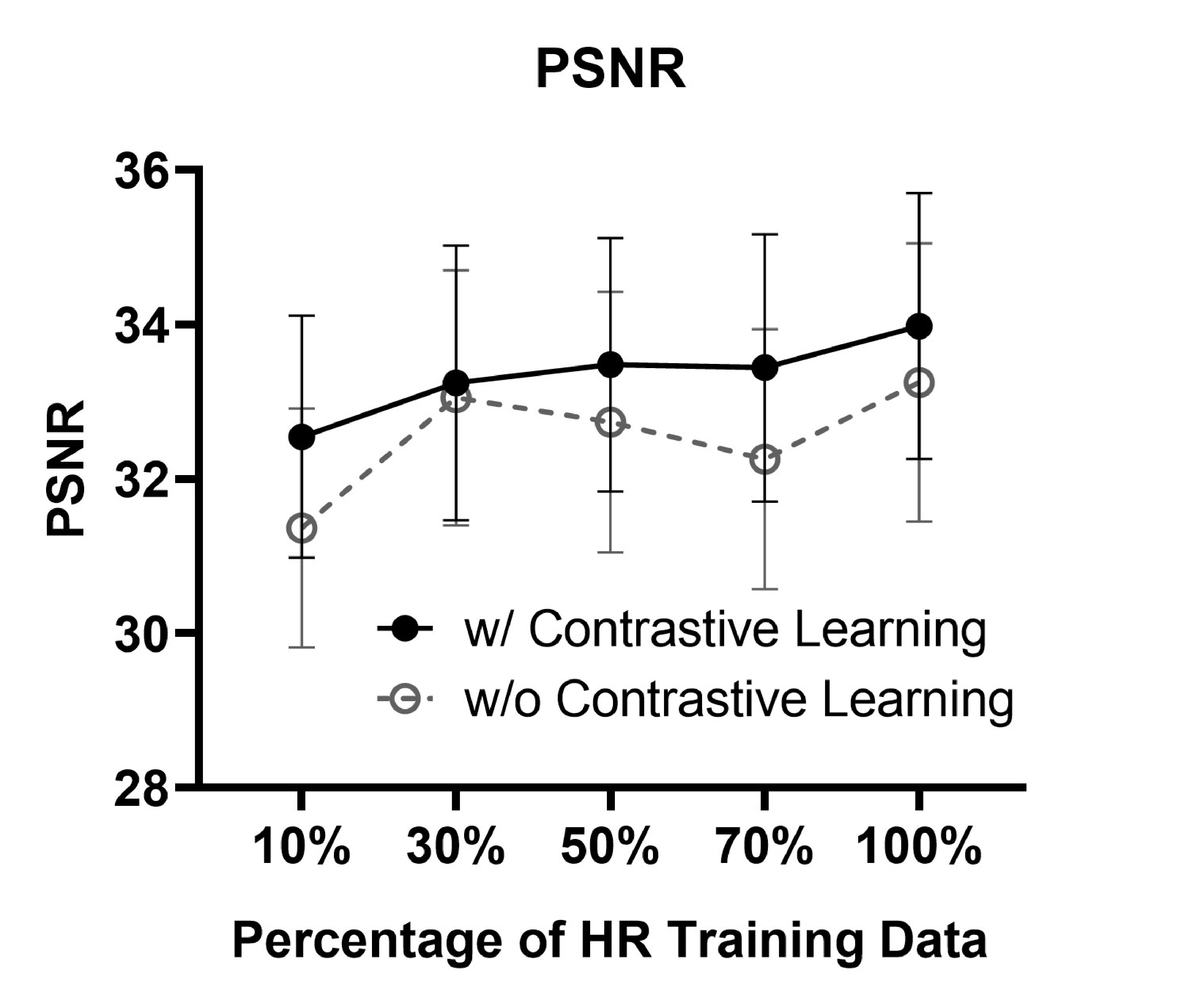}}
  \vspace{-0.1cm}
  \centerline{(b) }\medskip
\end{minipage}
\hfill
\vspace{-0.5cm}
\caption{Performance of models with various numbers of HR images. (a) and (b) are SSIM and PSNR results of ablation on contrastive loss for our method, respectively.}
\label{fig:res3}
\end{figure}

To assess the performance of our model under clinical conditions with limited HR training samples, we conducted experiments with varying numbers of HR images for training, and the corresponding test results are shown in Fig. \ref{fig:res3}. Additionally, we explored the impact of CL by comparing the proposed model with a conventional unsupervised model with identical architecture and removing contrastive
loss.

Fig. \ref{fig:res3} highlights that our proposed model with contrastive learning outperforms that without contrastive learning in nearly all cases. Furthermore, when the amount of accessible HR images diminishes, the SSIM/PSNR with only 30\% HR training data and contrastive learning (0.9144/33.2434 dB) has reached a comparable level with those with 100\% training data without contrastive learning (0.9231/33.2484 dB). We can also observe that contrastive learning is superior to its counterpart in the qualitative comparison shown in Fig. \ref{fig:res2}. As the number of training samples decreases, the generated SR images become increasingly blurry without contrastive loss. Also shown in the error maps, the model always achieves better accuracy and lower errors when contrastive loss is involved. Furthermore, the proposed method with 30\% HR images also outperforms the other state-of-the-art methods with 100\% HR images shown in Table \ref{metrics}, e.g., Blind-SR \cite{9999498} as the top of them achieves the SSIM/PSNR of 0.9132/33.0695 dB. 

\section{CONCLUSION}
\label{sec:majhead}

This study introduces a novel unpaired MRI SR network that adopts contrastive learning with an efficient strategy for constructing positive and negative sample pairs. The experimental results demonstrate that with a limited amount of HR training data, the performance of the proposed CL method is comparable to those of the unsupervised learning methods with a large number of training data. Furthermore, when only limited training data is provided, our method outperforms the unsupervised learning methods, demonstrating our method's superiority. Therefore, the proposed approach can be used in clinical settings where access to HR images is limited. Future research will explore more sophisticated sample pair construction and data augmentation strategies to enhance the model's inference capabilities.



\footnotesize
\bibliographystyle{IEEEbib}
\bibliography{strings,refs}

\end{document}